\documentclass[10pt,letterpaper]{article}
\usepackage[top=0.85in,left=2.75in,footskip=0.75in,marginparwidth=2in]{geometry}

\usepackage[utf8]{inputenc}

\usepackage{cite}

\usepackage{nameref,hyperref}

\usepackage[right]{lineno}

\usepackage{microtype}
\DisableLigatures[f]{encoding = *, family = * }

\raggedright
\usepackage{changepage}

\usepackage[aboveskip=1pt,labelfont=bf,labelsep=period,singlelinecheck=off]{caption}

\makeatletter
\renewcommand{\@biblabel}[1]{\quad#1.}
\makeatother

\usepackage{lastpage,fancyhdr,graphicx}
\usepackage{epstopdf}
\fancyheadoffset[L]{2.25in}
\fancyfootoffset[L]{2.25in}

\usepackage{color}

\definecolor{Gray}{gray}{.25}

\usepackage{graphicx}

\usepackage{sidecap}

\usepackage{lettrine}

\usepackage{wrapfig}
\usepackage[pscoord]{eso-pic}
\usepackage[fulladjust]{marginnote}
\reversemarginpar

\begin{document}
\vspace*{0.35in}

\newcommand{\be}{\begin{equation}}
\newcommand{\ee}{\end{equation}}
\newcommand{\Equat}[1]{Equation (\ref{eq:#1})}

\begin{flushleft}
{\Large
\textbf\newline{Principles of Information Storage in Small-Molecule Mixtures}
}
\newline
\\
Jacob K. Rosenstein\textsuperscript{1,*},
Christopher Rose\textsuperscript{1},
Sherief Reda\textsuperscript{1},
Peter M. Weber\textsuperscript{2},
Eunsuk Kim\textsuperscript{2},
Jason Sello\textsuperscript{2},
Joseph Geiser\textsuperscript{2},
Eamonn Kennedy\textsuperscript{1},
Christopher Arcadia\textsuperscript{1},
Amanda Dombroski\textsuperscript{2},
Kady Oakley\textsuperscript{2},
Shui Ling Chen\textsuperscript{2},
Hokchhay Tann\textsuperscript{1},
and Brenda M. Rubenstein\textsuperscript{2}
\\
\bigskip
\bf{1} School of Engineering, Brown University, Providence, RI 02912
\\
\bf{2} Department of Chemistry, Brown University, Providence, RI 02912
\\
\bigskip
* jacob\_rosenstein@brown.edu

\end{flushleft}

\section*{Abstract}
Molecular data systems have the potential to store information at dramatically higher density than existing electronic media. Some of the first experimental demonstrations of this idea have used DNA, but nature also uses a wide diversity of smaller non-polymeric molecules to preserve, process, and transmit information. In this paper, we present a general framework for quantifying chemical memory, which is not limited to polymers and extends to mixtures of molecules of all types. We show that the theoretical limit for molecular information is two orders of magnitude denser by mass than DNA, although this comes with different practical constraints on total capacity. We experimentally demonstrate kilobyte-scale information storage in mixtures of small synthetic molecules, and we consider some of the new perspectives that will be necessary to harness the information capacity available from the vast non-genomic chemical space.


\section*{Introduction}

\lettrine{A}{n} ever-increasing worldwide demand for digital data systems, alongside a looming slowdown of semiconductor technology scaling, has led to growing interest in molecular-scale platforms for information storage and computing. 
There have been several interesting demonstrations using DNA sequences to store abstract digital data, offering a path towards extremely dense archival information storage \cite{Zhirnov2016, Church2012}. Using tools developed for modern genomics, researchers have synthesized complex pools of oligomers representing hundreds of megabytes of text, images, videos, and other media files, and retrieved the data using commercial high-throughput sequencing instruments \cite{Church2012, Organick2018, Goldman2013, Grass2015, Erlich2017, Blawat2016}.

Other molecular information demonstrations have shown that a molecule could serve as a secret input to a chemical hash function \cite{Sarkar2016,Boukis2018}, and that two-dimensional arrays containing single compounds per grid position can encode digital data by photochemical or electrochemical means \cite{Green2007,Liu1990,Thomas2009}.
However, beyond examples describing polymers \cite{Colquhoun2014,Laure2016,Organick2018} or single molecules, the information capacity of molecular systems can be less intuitive. Given practical polymer synthesis constraints, can many small molecules store as much information as one macromolecule?

There are naturally two extremes of chemical information representations, with a continuum of possibilities between them. At one extreme, a single complex macromolecule can be synthesized such that its substructures (monomers) represent abstract data \cite{Colquhoun2014}. In the macromolecule regime, the challenge lies in the reliability and precision needed to synthesize and analyze such a large and complex molecule. At the other extreme, data could be spread across many simpler compounds, but here the challenge lies in precisely managing large diverse collections of molecules.

\begin{wrapfigure}{l}{110mm}
\includegraphics[width=110mm]{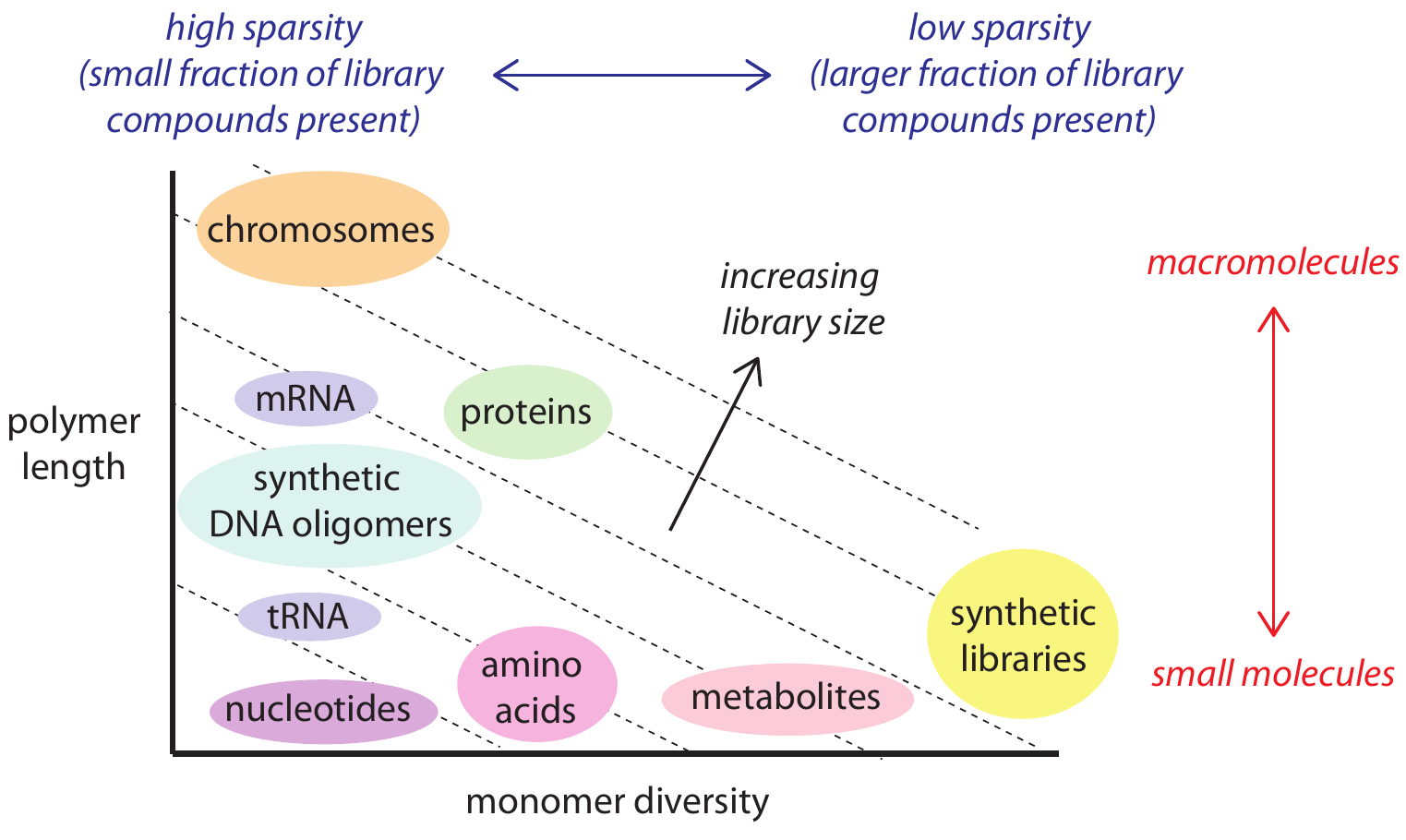}
\captionsetup{labelformat=empty}
\caption{}
\label{fig:biological_sparsity}
\end{wrapfigure}

Clearly mixtures of small molecules can represent and transfer information, as biology demonstrates with RNA, neurotransmitters, and metabolites (Fig.~\ref{fig:biological_sparsity}). Unfortunately, tools do not exist to quantify all of these types of information, hampering efforts to leverage them in synthetic biology \cite{Way2014} and synthetic data representations. 

\begin{wrapfigure}{l}{75mm}
\includegraphics[width=75mm]{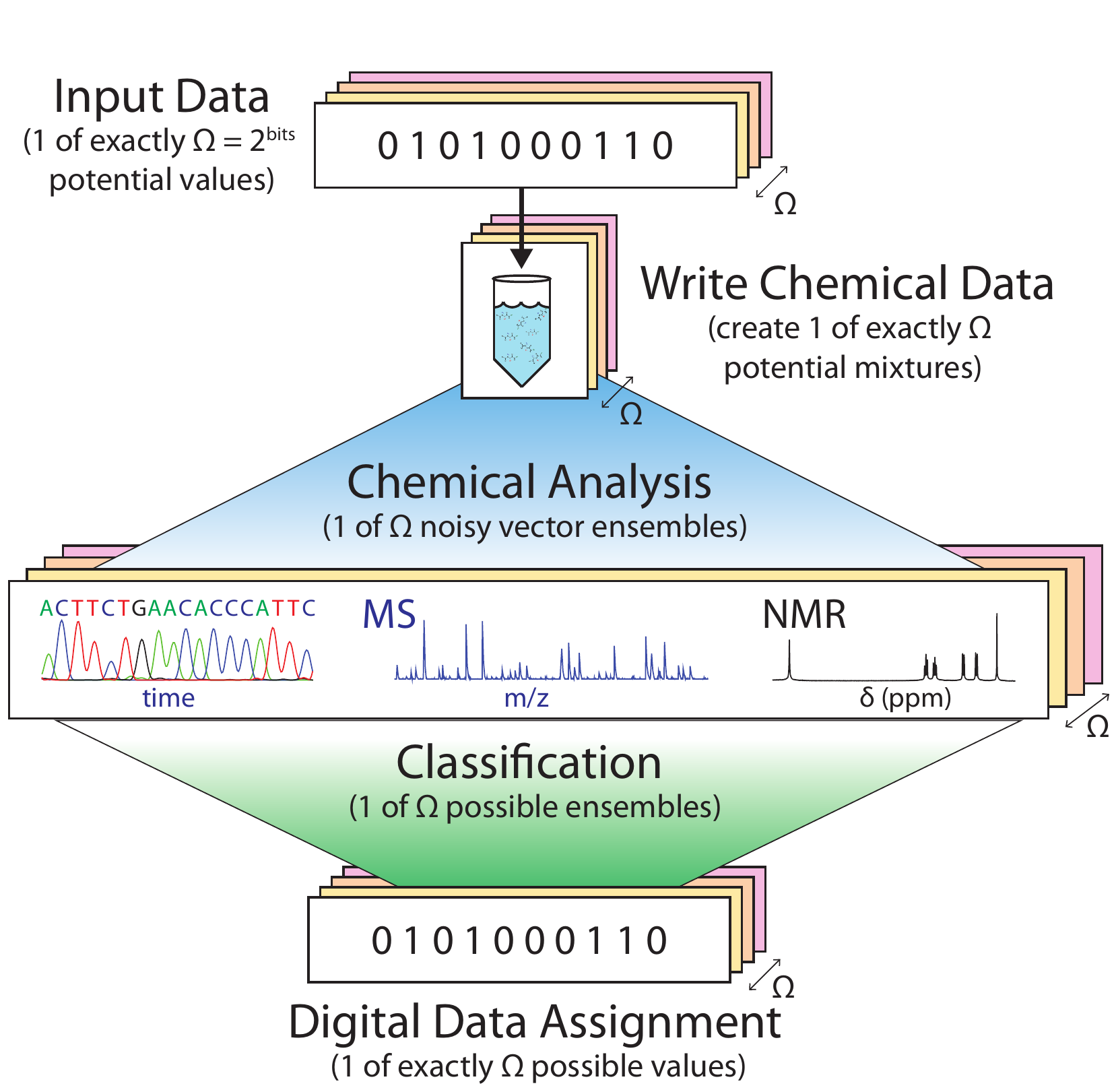}
\captionsetup{labelformat=empty}
\caption{}
\label{fig:analysis_space}
\end{wrapfigure}

In this paper, we present a general theory of information storage in molecules and in mixtures of molecules. This theory includes ordered polymers, while providing a unified description for other classes of molecules as well. This concept of molecular information is applicable to many different chemistries; the encoded data can be `read' using a variety of analysis techniques including mass spectrometry, sequencing, chromatography, or spectroscopy, as illustrated in Figure \ref{fig:analysis_space}. 

By introducing a more generalized framework for quantifying molecular information, we are optimistic that many new classes of molecular storage media will be developed, with valuable properties including even higher information density than DNA, beyond-biological chemical properties, and new dimensions for high speed chemical computing paradigms. Although few chemistries are as mature as those available for DNA, we show that diversified small-molecule approaches have intrinsic capacities for gigabyte-scale data storage. In addition to new experimental \cite{Kennedy_metabolites,Arcadia2018,ugi_unpublished} and theoretical tools for interrogating heterogeneous mixtures of molecules, this new perspective may also contribute to new ways of quantifying the information contained in the chemical states of living systems.

\marginpar{
\vspace{.7cm} 
\color{Gray} 
\textbf{Figure \ref{fig:analysis_space}.} 
Information is coded into a mixture of molecules from a predetermined library of possible chemicals. Reading a chemical memory corresponds to classifying it as one of exactly $\Omega$ values. Depending on the molecular library, any analysis technique which helps to differentiate among mixtures can be used. The shapes of the analysis vectors will be different from the shape of the data, but the number of possible states ($\Omega$) is finite, and will be the same at every stage.
}

\clearpage
\newpage

\marginpar{
\vspace{.7cm} 
\color{Gray} 
\textbf{Figure \ref{fig:biological_sparsity}.} 
Biological systems make use of both macromolecules and small molecules for information representations. Whereas long-term storage is encoded in ordered macromolecules (DNA), smaller and more chemically-diverse proteins and metabolites also represent large aggregate amounts of information that describe the working state of an organism.
}
~\newline
~\newline
~\newline
~\newline
~\newline
~\newline
~\newline
~\newline
~\newline
~\newline
~\newline
~\newline
~\newline
~\newline
~\newline
~\newline
~\newline
~\newline

\section{Foundations of Molecular Information Capacity}
Information is a measure of  {\em improbability}. If more potential states are available to a given system, it becomes less likely that one particular state will be realized. The information capacity of a system accounts for the number of possible states as well as the likelihood of confusing one state for another. If a chemical system has $\Omega$ identifiable states, then its information capacity ($C$, in bits) has an upper bound of
\begin{equation}
    C \le \log_2 \Omega.
\end{equation}

If we consider each molecule to be defined only by its chemical identity, we can quantify the amount of information represented in a chemical mixture by answering the following simple questions: (1) What is the set of unique molecules which could be present? (2) Which of these unique molecules are present? (3) How many copies of each unique molecule are present?


\subsection{Ordered Polymers}
To begin, consider linear polymers such as nucleic acids or proteins. Nucleic acids have four canonical bases, so the number of possible $N$-monomer strands is $M=4^N$. If only one of the $M$ molecules
may be present, then $\Omega=M$ and the identity of the molecule represents $2N$ bits. Similarly, proteins with $N$ monomers drawn from an alphabet of 20 amino acids carry $\log_2 20^N \approx 4.3N$ bits. The information capacity of a single polymer molecule is therefore expressed as
\be
\label{eq:Cmol}
C
\le
\log_2 M
=
N \log_2 B,
\ee
where $B$ is the number of different monomers. This result will be familiar to many readers.

Although it is often true that information is mapped independently onto substructures (monomers) within a molecule, it is equally true to say that it is actually the identity of the {\em whole} molecule which holds $\log_2M$ bits. (If one nucleotide changes, it is an entirely different molecule!) This concept is important for generalizing theories of information storage to more diverse non-polymeric molecules.

\subsection{Unordered Molecular Mixtures}
Now, consider an unordered mixture of up to $Q$ molecules. If exactly $Q$ molecules are drawn from a library of size $M$ (with potential duplication), then the total number of possible combinations is ${{M+Q-1}\choose{M-1}}$ \cite{Feller1960}.  If between $0$ and $Q$ molecules may be selected, then we have
\be
\label{eq:Omega}
\Omega
=
\sum_{q=0}^{Q}
{{M+q-1}\choose{M-1}}
=
\frac{Q+1}{M}
{{M+Q}\choose{M-1}}.
\ee
The capacity of the system is therefore
\be
\label{eq:Q}
C_1(M,Q)
\le
\log_2 \left [
  \frac{Q+1}{M}
{{M+Q}\choose{M-1}}
\right ].
\ee
If we do not allow duplication among the $Q$ selections, then
\be
\Omega
=
\sum_{q=0}^Q
{{M}\choose{q}},
\ee
so that the capacity is
\be
\label{eq:Qunique}
C_{2}(M,Q)
\le
\log_2 \left [
\sum_{q=0}^Q
{{M}\choose{q}}
\right ].
\ee
When all molecules may be present ($Q=M$) without duplication, this capacity becomes
\be
\label{eq:presence}
C_{2}(M,M)
\le
\log_2 \left [
\sum_{q=0}^M
{{M}\choose{q}}
\right ]
=
M \log_2 2,
\ee
which is simply $M$ bits.

\begin{wrapfigure}[10]{l}{75mm}
\includegraphics[width=75mm]{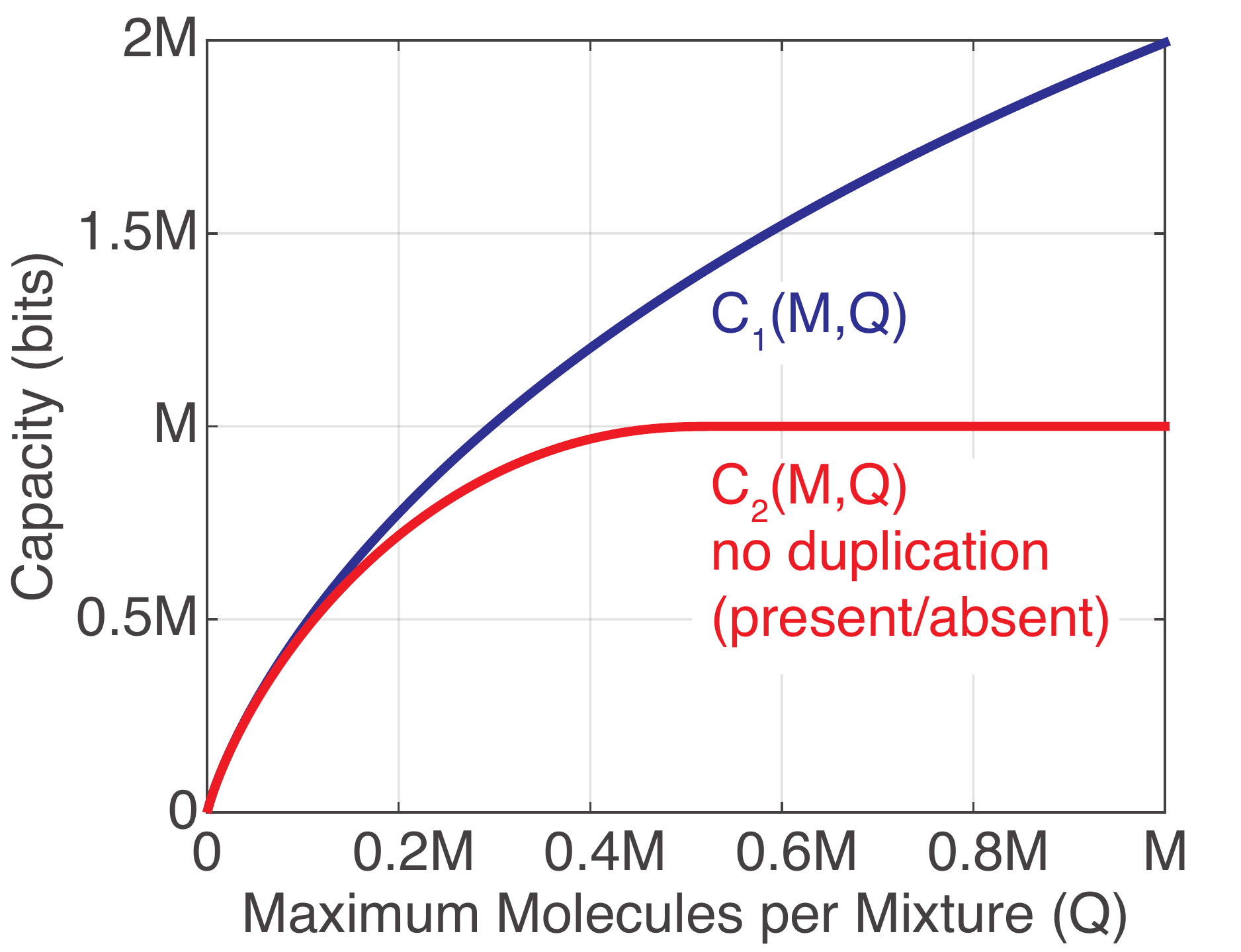}
\captionsetup{labelformat=empty}
\caption{}
\label{fig:memory}
\end{wrapfigure}

\marginpar{
\vspace{.7cm}
\color{Gray}
\textbf{Figure \ref{fig:memory}.}
Information capacity of a mixture as a function of the maximum number of molecules present (Q), from a library of $M$ molecules. If duplication carries no information, the capacity asymptotically approaches $C_2=M$ bits.
}
~\newline
~\newline
~\newline
~\newline
~\newline
~\newline
~\newline
~\newline
~\newline
~\newline
~\newline
~\newline
~\newline
~\newline
~\newline

It is worthwhile to note that $C_1$ is the larger of these capacities and provides an upper bound on \emph{all} memory schemes in unordered mixtures. However, making use of $C_1$ requires that we know the exact concentration (count) of each unique molecule. $C_2$ is the reduced capacity when duplication carries no information, which is also equivalent to classifying each unique molecule as simply ``absent'' or ``present'' above some concentration threshold. Representative curves are shown in Figure~\ref{fig:memory}. Without duplication, there are diminishing returns in information capacity as $Q$ approaches $M$.

In practical implementations of molecular memory, it is likely that many copies of each unique molecule will be present in a mixture. Rather than counting molecules, it may be more reasonable to specify that each of the $M$ molecules may exist at one of $L$ distinguishable concentrations. In this case, the capacity becomes
\begin{equation}
    C_{3}(M,L) \le C_{2}(M,M) \times \log_2 L = M \log_2 L,
    \label{eq:presence_concentration}
\end{equation}
which reduces to Equation~\ref{eq:presence} when $L$ = 2. 
Equation \ref{eq:presence_concentration} also applies when there are $L$ potential states of each of the $M$ library molecules, which may include chemical modifications or electronic, vibrational, or rotational states. It is important to note that $L$ is the number of {\em states}, not the number of dimensions. To reach this upper bound, each molecule's $L$ states must be independent. If the states only describe ensembles, the capacity multiplier will be less than $\log_2 L$.

\section{Molecular Data Addressing}

In an unordered mixture, all combinations (states) are equally valid, but there are practical advantages to re-introducing some ordering and hierarchy that will correspond to concepts of `addressing' within the data. The choice of chemical addressing scheme can have a large impact on the information density, the total capacity, and possibilities for random access.

\subsection{Spatial Addressing}

The most trivial form of addressing is spatial separation. Storing information across a set of independent chemical pools (such as in standard microwell plates) increases capacity linearly with the number of independent wells ($W$). Importantly, since wells are physically separated, the same library of $M$ potential molecules can be re-used in each well. In the limit of very small $Q$, spatial addressing also describes existing chemical microarrays \cite{Schena1995,Schirwitz2012} or two-dimensional molecular memory \cite{Green2007,Liu1990}.

\subsection{Sparse Data Mixtures and Address-Payload Coding}

Another valuable concept involves the subdivision of $M$ library molecules into groups of size $S$, and production of sparse mixtures which contain exactly one molecule from each subgroup. A mixture with sparsity $S$ will thus contain $M/S$ molecules. Since each molecule represents an exclusive choice among $S$ possibilities, the total capacity is 
\begin{equation}
    C_4(M,S) \le \frac{M}{S} \log_2 S,
    \label{eq:sparsecapacity}
\end{equation}
which is less than both $C_1$ and $C_2$.

\begin{wrapfigure}[10]{l}{75mm}
\includegraphics[width=130mm]{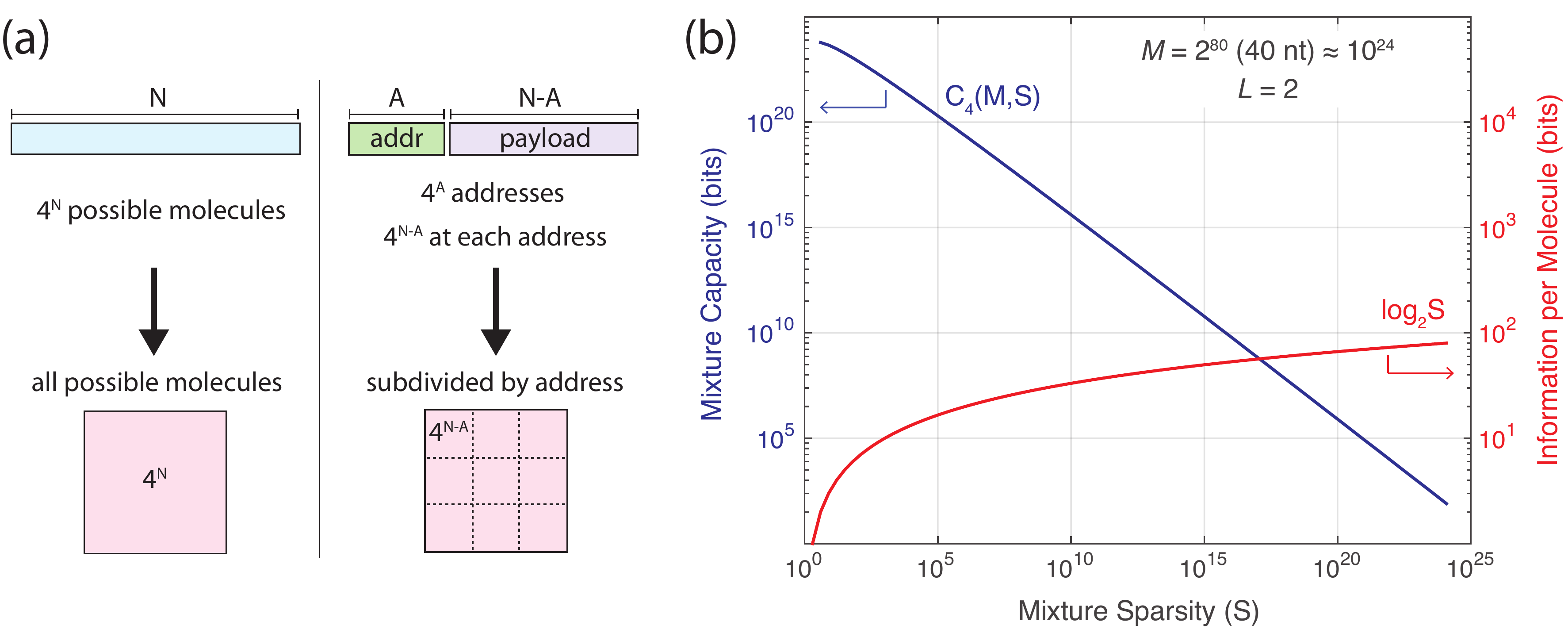}
\captionsetup{labelformat=empty}
\caption{}
\label{fig:addressing_and_sparsity}
\end{wrapfigure}

We note that the sparse mixture described by \Equat{sparsecapacity} is identical to an address-payload \cite{Bornholt2016} DNA data representation, as shown in Figure~\ref{fig:addressing_and_sparsity}a. By assigning $A$ positions in the sequence as an `address' and the remaining $N-A$ positions as a `payload,' the library of $M=4^N$ sequences has been subdivided using sparsity $S=4^{N-A}$, and exactly one sequence is included from each of the $4^A$ addresses. In DNA memory, this can be a productive strategy given constraints on DNA synthesis length \cite{Church2012,Organick2018}.

\newpage

\marginpar{
\vspace{.7cm}
\color{Gray}
\textbf{Figure \ref{fig:addressing_and_sparsity}.}
(a) Mixture sparsity and DNA address-payload representations in molecular datasets. By requiring that each mixture contains exactly one molecule per address space, one can balance the benefits of smaller data mixtures against a reduced total information capacity for a given library. \\(b) Increasing mixture sparsity ($S$) produces mixtures with fewer molecules, and confers more information per unique molecule present. However, the maximum total capacity corresponds to the densest mixtures because the information per molecule scales only logarithmically with the sparsity.
}
~\newline
~\newline
~\newline
~\newline
~\newline
~\newline
~\newline
~\newline
~\newline
~\newline
~\newline
~\newline
~\newline
~\newline
~\newline
~\newline
~\newline

Enforced sparsity reduces the number of valid mixture states ($\Omega$), by disallowing mixtures which contain more than one molecule from the same address space. 
The information conveyed per molecule increases, but the overall mixture capacity is reduced. Non-polymeric chemical memories may similarly benefit from sparse representations, as increased sparsity can imply synthesizing fewer molecules and analyzing simpler mixtures.

\subsection{Capacity Implications}

\begin{wrapfigure}[10]{l}{100mm}
\includegraphics[width=100mm]{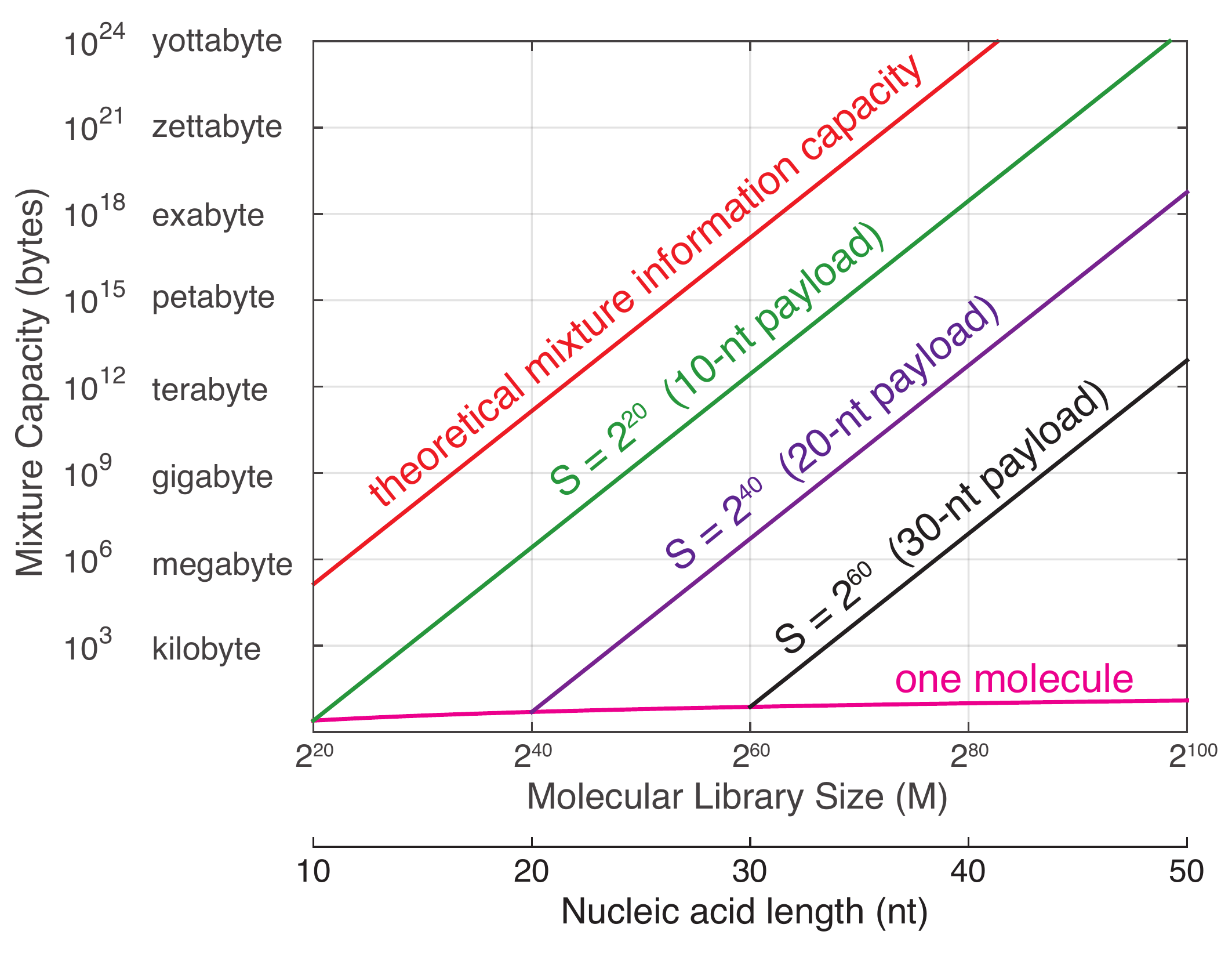}
\captionsetup{labelformat=empty}
\caption{}
\label{fig:capacityplot}
\end{wrapfigure}

These mixture capacity analyses have some simple but perhaps nonintuitive implications. As shown in Figure~\ref{fig:addressing_and_sparsity}b, the maximum per-molecule information density occurs for maximum sparsity ($S=M$), but the maximum total mixture capacity is achieved with the minimum sparsity ($S=1$). In other words, for a fixed-size library, the maximum mixture capacity is reached when each molecule represents only an address, with no payload! In theory, a library consisting of short DNA oligomers of length $N=40$ could either be used to select one molecule conveying 80 bits, or it could be used to create one unordered molecular mixture which represents 151 zettabytes ($151 \times 10^{21}$ bytes) of data, which is on the scale of all of the digital information produced in the entire world per year (Figure~\ref{fig:capacityplot}) \cite{Cisco2016,Zhirnov2016}. If only single copies of each molecule were present (or absent), this hypothetical data set would weigh only a few pounds. In practice, such experiments are of course limited by chemical synthesis throughput.

However impractical, this thought experiment underlines the fact that while long DNA synthesis and long-read sequencing are real bottlenecks for some biological applications \cite{Jain2018,Kosuri2014}, mixtures of short polymers would be more than capable of representing any fathomable amount of digital data. Scaling DNA data storage should focus on increasing throughput, rather than length \cite{Church2018}. This perspective also suggests that many other families of molecular libraries should be compatible with gigabyte-scale information mixtures, even when lacking the exponential library scaling of long polymers.

\newpage

\marginpar{
\vspace{.7cm}
\color{Gray}
\textbf{Figure \ref{fig:capacityplot}.}
Information capacity of molecular mixtures. Plotting the capacity for several different sparsities shows the potential of complex chemical mixtures for large-scale data storage. The capacity of one molecule scales logarithmically with the library size ($M$), but the capacity of a mixture scales linearly. In theory, all of the digitized information produced in the world each year could be stored in one unordered mixture of short 40-nt DNA molecules.
}
~\newline
~\newline
~\newline
~\newline
~\newline
~\newline
~\newline
~\newline
~\newline
~\newline
~\newline
~\newline
~\newline
~\newline
~\newline
~\newline
~\newline
~\newline
~\newline

\subsection{Energy Constraints of Molecular Memory}

Any implementation of molecular memory will face constraints in both synthesizing the library and creating the data mixtures. Given the tradeoffs between library size ($M$), mixture size ($Q$), and number of independent mixtures ($W$), what would constitute an optimal design? It seems worthwhile to consider the costs of representing the same information in different configurations. For a mixture of polymers, if we assume the marginal energy per monomer incorporation is $\epsilon$, then $W$ mixtures of $Q$ unique molecules with length $N$ would require a total energy of
\begin{equation}
{\cal E} = \epsilon WQN.
\end{equation}
For $W$ independent mixtures, we can rewrite \Equat{sparsecapacity} as
\begin{equation}
C \le WQ \log_2 \frac{M}{Q} = WQ \Big( \log_2 M - \log_2 \frac{M}{S} \Big),
\end{equation}
from which we can see that for very sparse mixtures (including single molecules), the second term is negligible. Substituting $M=B^N$, we can solve for the energy per bit (${\cal E}_b$)
\begin{equation}
{\cal E}_b = \frac{{\cal E}}{C} \approx \frac{\epsilon}{\log_2 B},
\end{equation}
which suggests that for very sparse mixtures of polymers, there are energy benefits from increasing monomer diversity ($B$), although the scaling is sublinear.

On the other hand, for dense binary mixtures (large $Q$) which may contain many unique compounds, recall from \Equat{presence} and Figure~\ref{fig:memory} that $C \approx M$ per well. In many datasets, we can also approximate $Q \approx M/2$. Thus,
\begin{equation}
{\cal E}_b = \frac{{\cal E}}{C} \approx \frac{\epsilon N}{2},
\end{equation}
which implies that the optimal strategy is to produce mixtures using the simplest molecules (smallest $N$) capable of yielding mixtures with the desired capacity.

Across multiple dense mixtures one can see that there will be many duplicated syntheses. If the entire library is synthesized ahead of time, the synthesis cost will be amortized, and the energy constraint may be better described by a physical mixing or fluid handling cost ($\gamma$)
\begin{equation}
{\cal E} = \gamma WQ \approx \gamma W\frac{M}{2} = \gamma \frac{C}{2}
\end{equation}
and thus the energy per bit is a constant
\begin{equation}
{\cal E}_b = \frac{{\cal E}}{C} \approx \frac{\gamma}{2},
\end{equation}
which unfortunately reveals no obvious opportunity for the optimization of write costs for dense molecular mixtures.

To minimize the sizes of both the pre-synthesized library and the array of mixtures, it may be reasonable to optimize for $\min (M+W)$ while maintaining $C=MW$. Geometrically this is a minimum perimeter problem, satisfied by
\begin{equation}
W \approx M \approx \sqrt{C},
\end{equation}
which is interesting in its implication that, for dense mixtures, one optimum occurs when the data mixtures' spatial diversity and molecular diversity are similar.

\section{Diversified Small-Molecule Memory}

A simple summary of the preceding analysis is that a library of $M$ unique molecules can produce a binary mixture representing as few as $\log_{2}M$ bits and as many as $M$ bits of information (\Equat{presence}). There are at least $10^{5}$ known biological metabolites \cite{MS_metabolites_database,Kennedy_metabolites}), and far more synthetically feasible small molecules.

Even among small organic molecules, there are potentially more than $10^{60}$ unique compounds \cite{Bohacek1996}, and within this vast space, there may be many potential targets for megabyte- and gigabyte-scale small-molecule libraries. 

Combinatorial chemistries are regularly used in pharmaceutical pipelines to explore the space of potential drug candidates \cite{Gerry2018,Schreiber2000}. One of the most scalable strategies for generating functional group diversity is using multicomponent reactions (MCRs)\cite{Malinakova2015}. MCRs, which include the Hantzsch, Biginelli, Passerini, and Ugi reactions, are chemical transformations in which three or more reactants combine, largely independent of the order in which they are added, to form a single, multicomponent product. Because there are hundreds to thousands of different commercially-available possibilities for each reactant, MCRs can generate extremely large libraries. For example, recently reported five-dimensional Ugi-Petasis reactions can theoretically span a chemical space of at least  $1000\times200\times500\times1000\times1000=10^{14}$ molecules \cite{Portlock2003, Domling2006}. Perhaps the largest small molecule library reported to date was produced using a single split-pool synthesis and contained more than two million different compounds \cite{Tan1999}. Pharmaceutical companies routinely synthesize and screen millions of compounds \cite{Schreiber2000}, and as of 2015, the digital repository PubChem contained more than 60 million distinct chemical structures \cite{PubChem}. 

In total, the number of unique compounds synthesized worldwide to date is likely in the billions, yet this is still only a small fraction of the theoretical chemical space \cite{Borrel2018}. Even when restricted to only 17 or fewer atoms, a recent simulated enumeration of chemically stable and synthetically feasible organic molecules predicted more than 166 billion possible small organic molecules \cite{Ruddigkeit2013}. Some of the unrealized molecules contain chiral centers and ring systems that remain a challenge to produce using diversity-oriented techniques \cite{Schreiber2000}. Yet even with these synthetic challenges, there remains ample room for the design and discovery of new classes of molecules for information systems \cite{Frenkel2010}.

One serious challenge with molecular memory in unexplored chemical spaces is that readout options are far less mature than those for DNA. However, it is not necessary to have a single unambiguous measurement of each molecule present; the goal is only to recover the encoded information, which can be designed to tolerate some chemical ambiguity and errors.

\section{Reading Molecular Memories} 

\subsection{Detection Signal Spaces}

Depending on the chemical library, sequencing, mass spectrometry, optical spectroscopy, NMR, or chromatography may all be leveraged to analyze molecular mixtures, and thereby read the data. The detection signal space is typically larger than the chemical mixture space, but the critical goal is simply to uniquely identify each of the $\Omega$ potential mixtures, as illustrated in Figure \ref{fig:analysis_space}.

It is advantageous when the detection signal space maps directly to the molecules in the library. For example, DNA sequencing schemes are generally designed to produce fluorescence or pH time series which correspond to nucleic acid sequences \cite{Mardis2017}. Yet  this one-to-one correspondence is not mandatory, and users of nanopore sequencing platforms have shown that chemical structure can be reliably decoded from extremely complex signals if the signals are repeatable and training datasets are available \cite{Rang2018}. Statistical approaches which identify correlated variables and reduce dimensionality \cite{Aeron2010} 
will often be required to disambiguate signals from data mixtures of non-genomic compounds. For example, infrared absorbance and Raman spectroscopy enable highly specific fingerprinting of molecules within complex mixtures, using rapidly improving optical sources and statistical tools \cite{Schliesser2012}. In Section~\ref{section_experimental}, we will introduce a methodology which uses mass spectrometry (Fig.~\ref{fig:experimental_ibex}).

\subsection{Capacity Under Detection Limits}

All of the information capacity expressions thus far have been upper bounds, which are only achievable  if there are no errors. As we will see in our experiments, detection errors that mistake one mixture for another are likely to occur. However, since these errors are probabilistic, there are many ways to encode data so that retrieval is asymptotically error-free. Each error correction scheme comes with a penalty of reduced total capacity 
\cite{Cover2012,Polyanskiy2010}. 

The upper limit for the capacity of a memory system can be described by its `confusion matrix,' which quantifies the probabilities of mistaking one of the $\Omega$ mixtures for another. 
If we let $P_{ii} = P_c$ and assume worst case equiprobable confusion ($P_{i\ne j} = \frac{1-P_c}{\Omega - 1}$), then we have
\be
\label{eq:memcapacity}
C' = \log_2\Omega +   P_c\log_2 P_c 
 +   (1-P_c) \log_2 \left ( \frac{1-P_c}{\Omega-1}  \right ).
\ee
If there is never any confusion ($P_c =1$), the capacity reaches its maximum of $\log_2 \Omega$ bits. If $\Omega$ is large, we can approximate
\begin{equation}
C'
\approx
P_c \log_2 \Omega - H_B(P_c),
\end{equation}
where $H_B(\cdot)$ is the binary entropy function \cite{Cover2012}. Thus, the information capacity scales almost linearly with the probability of correctly identifying the chemical state ($P_c$).

\subsection{Channel Coding and Error Correction}
\label{sec:ecc}
Implicit in the capacity expression (\Equat{memcapacity}) is the idea that we will need to tolerate some errors in identifying mixtures, while minimizing errors in the data assignments. It is well known that by spreading data across {\em sequences} of binary inputs (`codewords') of length $N_c$, the probability of errors after decoding can be made vanishingly small if the number of valid codewords $|c|$ satisfies 
\begin{equation}
\frac{\log_2 |c|}{N_c} < C',
\end{equation}
where $C'$ is the capacity of the system (in bits) which incorporates expected error rates. For example, to encode 10 bits of information using a library of $M=20$ molecules, we might designate only $|c|=2^{10}$ binary mixtures as `valid' out of the $\Omega=2^{20}$ mixtures which are possible. Since $|c|<\Omega$, channel coding can be thought of as another form of strategic sparsity, although it constrains the valid states in more sophisticated ways than limiting the number of molecules present. When analysis noise and errors result in an invalid mixture state, the decoder can classify it as the `nearest' valid codeword, by some metric. 

Successful DNA memory demonstrations have utilized Reed-Solomon codes and fountain codes \cite{Organick2018,Erlich2017}, which are robust error correcting codes (ECC), but can add significant complexity and capacity penalties.
Modern communications systems offer a number of practical methods for constructing near-capacity codes. One intriguing newer candidate for such applications is recent work on ``noise guessing''  \cite{Duffy2018a}, where a codebook is chosen (usually using known codes, but a random codebook is also possible), and upon detection, a finite series of maximum likelihood noise sequences are applied to the channel output sequentially. This new ``channel-centric'' method is both surprisingly efficient and capacity-achieving in the limit of large $N_c$.

\section{Experimental Demonstrations}
\label{section_experimental}

To explore physical implementations of these concepts, several experimental demonstrations were performed. Digital data was written into molecular mixtures using a programmable acoustic liquid handler (Labcyte Echo 550). Droplets from chemical libraries were deposited onto steel plates at 2.25 mm pitch, with 1536 mixture spots per plate. To recover the data, Fourier-transform ion cyclotron resonance (FT-ICR) mass spectrometry was used to analyze and estimate the chemical mixture in each spot (SolariX 7T, Bruker).

\newpage

\begin{wrapfigure}{l}{120mm}
\includegraphics[width=120mm]{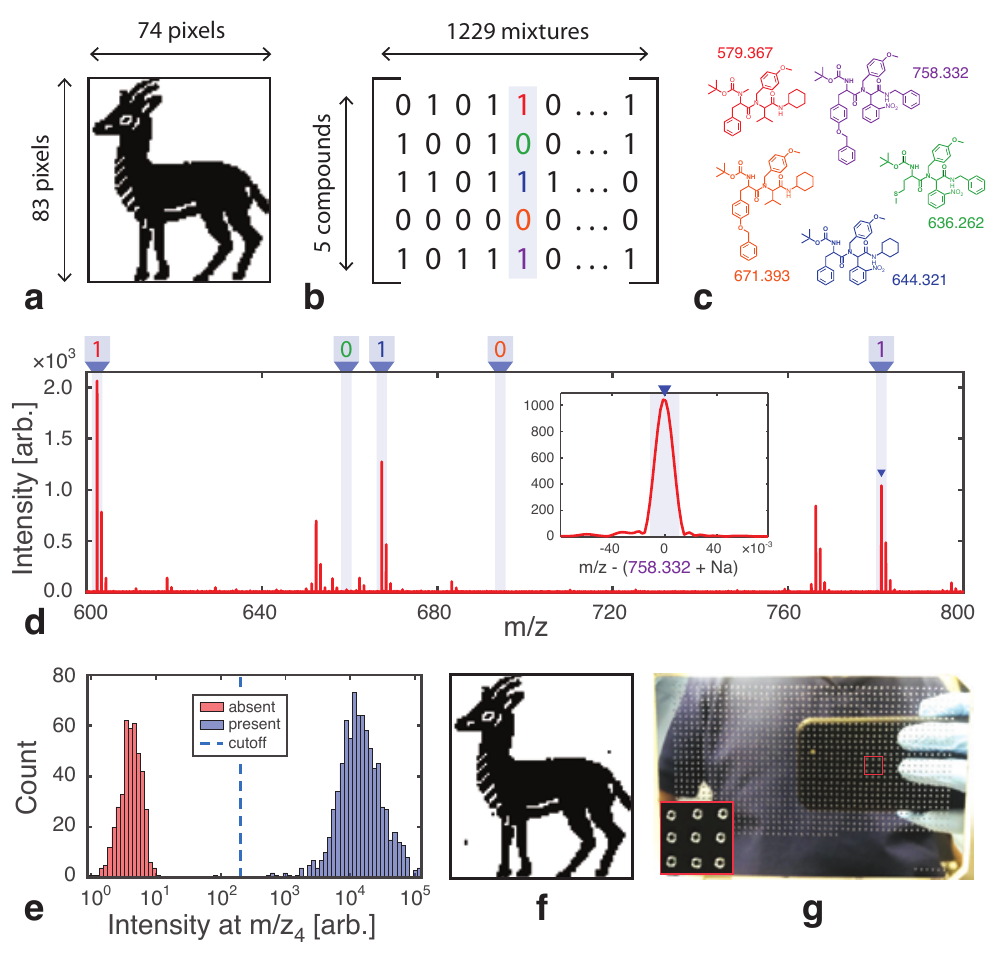}
\captionsetup{labelformat=empty}
\caption{}
\label{fig:experimental_ibex}
\end{wrapfigure}

Figure \ref{fig:experimental_ibex} illustrates one example of writing and reading a small digital image of an ibex from an Egyptian block print \cite{ibex}. A library of five small organic compounds (Fig.~\ref{fig:experimental_ibex}c) was synthesized, and mixtures were assembled in which each binary image pixel mapped onto the presence or absence of one compound in one mixture (as described by \Equat{presence}). To read back the data, the data was analyzed by mass spectrometry and the presence of each of the five library compounds was determined from the intensity of its primary sodiated ion. The digital image was recovered with 99.93\% accuracy.

\marginpar{
\vspace{.7cm} 
\color{Gray} 
\textbf{Figure \ref{fig:experimental_ibex}.} 
Experimental realization of information storage in small-molecule mixtures. (a) The dataset is a 6,142-pixel binary image of a Nubian ibex \cite{ibex}. (b) The data was mapped onto mixtures of five small organic compounds. (c) Chemical structures and masses of the five compounds. (d) A mass spectrum of one of the mixtures, with vertical lines denoting the masses corresponding to library compounds. This mixture represents the five bits `10101.' (e) A histogram of the measured sodiated peak intensities for one of the compounds shows a clear separation between the present (`1') and absent (`0') compounds. (f) These two distributions were were seperated with Fisher's linear discriminant, and the image was reconstructed with an error rate of 4/6142 = 0.065$\%$. (g) An image of the 1229 data mixtures, spotted on a steel plate for analysis by mass spectrometry (MS).
}

\clearpage
\newpage

\begin{wrapfigure}{l}{120mm}
\includegraphics[width=120mm]{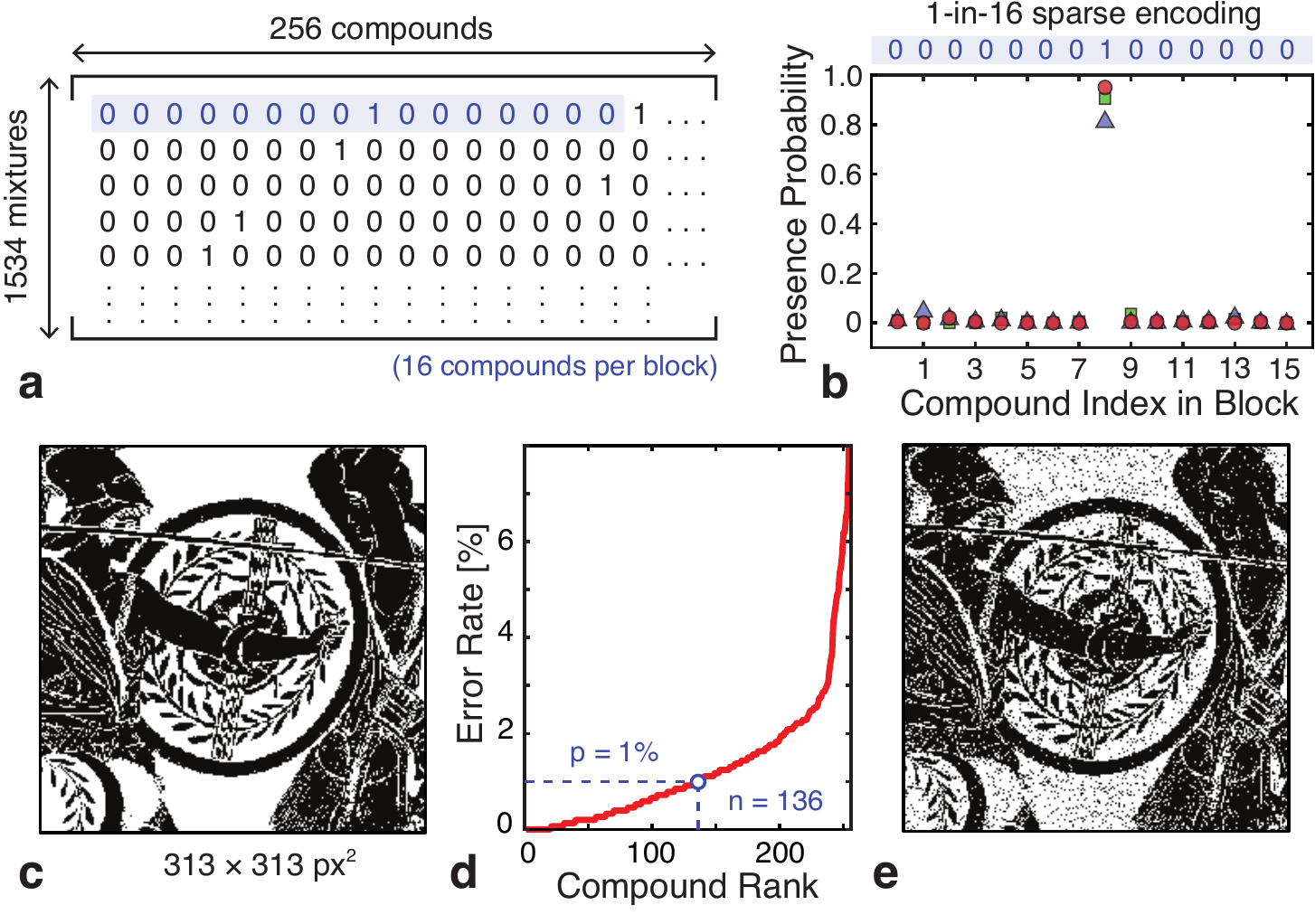}
\captionsetup{labelformat=empty}
\caption{}
\label{fig:amazonomachy}
\end{wrapfigure}

As a second demonstration (Fig.~\ref{fig:amazonomachy}), we experimentally implemented a sparse encoding scheme (described by \Equat{sparsecapacity} with $S$=16) to encode an image of Amazonomachy from a piece of Greek pottery \cite{amazonomachy}. A library of size $M$=256 was subdivided into 16 blocks, and groups of 4 binary pixels were mapped onto a one-hot selection of 1-from-16 compounds to include in the mixture (Fig.~\ref{fig:amazonomachy}a). To encode the 97,969 bit image, 1534 mixtures were created, which each contained 16 molecules and represented 16$\times\log_2$16=64 bits/mixture. Thanks to the sparsity of the mixtures, each present molecule encodes 4 bits of information. 

The Amazonomachy mixtures were similarly analyzed by mass spectrometry. A regression predicted which compound in each block was present with the highest signal-to-noise ratio (Fig.~\ref{fig:amazonomachy}b). From this analysis, 136 out of the 256 compounds yielded $<$1\% raw presence/absence error (Fig.~\ref{fig:amazonomachy}d). After decoding, the recovered digital image was 94.6$\%$ accurate (Fig.~\ref{fig:amazonomachy}e).
%

\marginpar{
\vspace{.7cm} 
\color{Gray} 
\textbf{Figure \ref{fig:amazonomachy}.} 
Experimental data storage in sparse molecular mixtures. (a) Here, data was encoded using a library of 256 small molecules at sparsity $S$=16 across 1534 mixtures. Groups of four pixels are mapped onto one-hot sequences of 16 compounds, such that each present molecule represents 4 bits of information. (b) The data is analyzed with mass spectrometry, and three example decoded blocks are shown with compound \#8 present (`1000'). (c) Using this scheme, a 97,969 pixel binary image was encoded depicting Amazonomachy from a piece of Greek pottery \cite{amazonomachy}. (d) Reading back the data using MS, 136 out of the 256 library compounds yielded $<$1\% raw error. (e) After decoding, the overall recovered image accuracy was 94.6$\%$.
}
~\newline
~\newline
~\newline
~\newline
~\newline
~\newline
~\newline
~\newline
~\newline
~\newline
~\newline
~\newline
~\newline
~\newline
~\newline
~\newline
~\newline
~\newline
~\newline
~\newline
~\newline
~\newline
~\newline

\section{Discussion}

By developing a formal theory of the information capacity of mixtures of molecules, we have shown how information can be represented by any chemical library. Regardless of the types of molecules, the identities and concentrations of molecules within a mixture can serve as atomic-scale representations of abstract digital data. We have demonstrated these ideas experimentally using several families of small molecules, including the demonstrations in Figure~\ref{fig:experimental_ibex} and Figure~\ref{fig:amazonomachy}, as well as other datasets using phenols \cite{Arcadia2018}, metabolites \cite{Kennedy_metabolites}, and multi-component reaction products \cite{ugi_unpublished}. These experiments have significant room for growth, using error correcting codes and expanded chemical libraries.

Although it is easier to conceptualize information storage within a single polymer, this perspective reminds us that single-molecule complexity and mixture complexity are complementary dimensions. The sparsity of a mixture relative to the available library size allows us to quantify the compromise between the challenges of both extremes. In scenarios when it is feasible to synthesize every compound from the library, denser mixtures provide higher total information capacity, even when the constituent molecules are polymers themselves.


Demonstrations of DNA data storage have exceeded 200 megabytes \cite{Organick2018}, but although this stretches today's synthesis capabilities, it represents a tiny fraction of the potential of molecular data storage. Organick \textit{et al.} synthesized 3.2 million unique $\approx$ 110-nt sequences; this is a mixture with a sparsity ($S$) of only one out of every $\approx10^{59}$ molecules from the library. As technologies for higher throughput synthesis evolve \cite{Church2018,Carlson2009}, even if they are accompanied by higher error rates, DNA memory still has tremendous room for growth.

In non-genomic chemical space, working within the assumptions that led to an estimate of $10^{60}$ drug-like small molecules \cite{Bohacek1996}, the selection of one 500 Da molecule could represent as much as log$_{2}10^{60} \approx$ 200 bits. To represent the same amount of information in DNA would require a molecule with a mass of 65,000 Da. Despite the practical limitations of this comparison, we can recognize opportunities for chemical information systems with up to two orders of magnitude lower mass than DNA, and with far greater chemical diversity.  



Modern information technology is moving towards a more unified vision of computation and memory, and fluid molecular mixtures offer an intriguing space for future generations of computing systems that take advantage of the natural complexity and intrinsic statistics of chemical systems \cite{Arcadia2018,Rose2018,Kennedy2018,Jiang2012,Soloveichik2010,Chen2014}. More precisely quantifying the information capacity of chemical mixtures represents an early step in this direction, and we anticipate that valuable scientific advances may come from using this lens to consider pathways within mixtures of reactive chemical libraries.



\section*{Acknowledgments}
This research was supported by funding from the Defense Advanced Research Projects Agency (DARPA W911NF-18-2-0031). The views, opinions and/or findings expressed are those of the authors and should not be interpreted as representing the official views or policies of the Department of Defense or the U.S. Government.

\nolinenumbers

\bibliography{JKR_mixtures}

\begin{thebibliography}{10}

\bibitem{Aeron2010}
S.~Aeron, V.~Saligrama, and M.~Zhao.
\newblock {Information Theoretic Bounds for Compressed Sensing}.
\newblock {\em IEEE Transactions on Information Theory}, 56(10):5111--5130,
  2010.

\bibitem{Arcadia2018}
C.~Arcadia, H.~Tann, A.~Dombroski, K.~Ferguson, S.~Chen, E.~Kim, C.~Rose,
  B.~Rubenstein, S.~Reda, and J.~K. Rosenstein.
\newblock {Parallelized Linear Classification with Volumetric Chemical
  Perceptrons}.
\newblock In {\em Proceedings of the IEEE Conference on Rebooting Computing
  (ICRC)}, 2018.

\bibitem{ugi_unpublished}
{Arcadia et al}.
\newblock {\em In preparation}.

\bibitem{Blawat2016}
M.~Blawat, K.~Gaedke, I.~H{\"{u}}tter, X.-M. Chen, B.~Turczyk, S.~Inverso,
  B.~W. Pruitt, and G.~M. Church.
\newblock {Forward Error Correction for DNA Data Storage}.
\newblock {\em Procedia Computer Science}, 80:1011--1022, 2016.

\bibitem{Bohacek1996}
R.~S. Bohacek, C.~McMartin, and W.~C. Guida.
\newblock {The art and practice of structure-based drug design: A molecular
  modeling perspective}.
\newblock {\em Medicinal Research Reviews}, 16(1):3--50, sep 1996.

\bibitem{Bornholt2016}
J.~Bornholt, R.~Lopez, D.~M. Carmean, L.~Ceze, G.~Seelig, and K.~Strauss.
\newblock {A DNA-Based Archival Storage System}.
\newblock In {\em Proceedings of the Twenty-First International Conference on
  Architectural Support for Programming Languages and Operating Systems},
  ASPLOS '16, pages 637--649, New York, NY, USA, 2016. ACM.

\bibitem{Borrel2018}
A.~Borrel, N.~C. Kleinstreuer, and D.~Fourches.
\newblock {Exploring drug space with ChemMaps . com}.
\newblock {\em Bioinformatics}, (1):1--3, 2018.

\bibitem{Boukis2018}
A.~C. Boukis, K.~Reiter, M.~Fr{\"{o}}lich, D.~Hofheinz, and M.~A.~R. Meier.
\newblock {Multicomponent reactions provide key molecules for secret
  communication}.
\newblock {\em Nature Communications}, 9(1):1439, 2018.

\bibitem{Carlson2009}
R.~Carlson.
\newblock {The changing economics of DNA synthesis}.
\newblock {\em Nature Biotechnology}, 27:1091, dec 2009.

\bibitem{Chen2014}
H.-L. Chen, D.~Doty, and D.~Soloveichik.
\newblock {Deterministic function computation with chemical reaction networks}.
\newblock {\em Natural Computing}, 13(4):517--534, 2014.

\bibitem{Church2018}
G.~M. Church.
\newblock {Enzymatic DNA synthesis for digital information storage}.
\newblock {\em bioRxiv}, 2018.

\bibitem{Church2012}
G.~M. Church, Y.~Gao, and S.~Kosuri.
\newblock {Next-Generation Digital Information Storage in DNA}.
\newblock {\em Science}, 337(6102):1628 LP -- 1628, sep 2012.

\bibitem{Cisco2016}
Cisco.
\newblock {The Zettabyte Era : Trends and Analysis}.
\newblock 2016.

\bibitem{Colquhoun2014}
H.~Colquhoun and J.-F. Lutz.
\newblock {Information-containing macromolecules}.
\newblock {\em Nature Chemistry}, 6:455, may 2014.

\bibitem{Cover2012}
T.~Cover and J.~A. Thomas.
\newblock {\em {Elements of information theory}}.
\newblock Wiley, 2nd editio edition, 2012.

\bibitem{Domling2006}
A.~D{\"{o}}mling.
\newblock {Recent Developments in Isocyanide Based Multicomponent Reactions in
  Applied Chemistry}.
\newblock {\em Chemical Reviews}, 106(1):17--89, jan 2006.

\bibitem{Duffy2018a}
K.~R. Duffy, J.~Li, and M.~Medard.
\newblock {Capacity-achieving decoding by guessing noise}.
\newblock {\em arXiv}, (2), 2018.

\bibitem{Erlich2017}
Y.~Erlich and D.~Zielinski.
\newblock {DNA Fountain enables a robust and efficient storage architecture}.
\newblock {\em Science}, 355(6328):950 LP -- 954, mar 2017.

\bibitem{Feller1960}
W.~Feller.
\newblock {\em {An Introduction to Probability Theory and Its Applications}}.
\newblock John Wiley and Sons, Inc, 1960.

\bibitem{Frenkel2010}
D.~Frenkel and B.~Smit.
\newblock {\em {Understanding molecular simulation: from algorithms to
  applications}}.
\newblock Elsevier, 2010.

\bibitem{Gerry2018}
C.~J. Gerry and S.~L. Schreiber.
\newblock {Chemical probes and drug leads from advances in synthetic planning
  and methodology}.
\newblock {\em Nature Reviews Drug Discovery}, 17:333, apr 2018.

\bibitem{Goldman2013}
N.~Goldman, P.~Bertone, S.~Chen, C.~Dessimoz, E.~M. LeProust, B.~Sipos, and
  E.~Birney.
\newblock {Towards practical, high-capacity, low-maintenance information
  storage in synthesized DNA}.
\newblock {\em Nature}, 494:77, jan 2013.

\bibitem{Grass2015}
R.~N. Grass, R.~Heckel, M.~Puddu, D.~Paunescu, and W.~J. Stark.
\newblock {Robust Chemical Preservation of Digital Information on DNA in Silica
  with Error-Correcting Codes}.
\newblock {\em Angewandte Chemie International Edition}, 54(8):2552--2555, feb
  2015.

\bibitem{Green2007}
J.~E. Green, J.~{Wook Choi}, A.~Boukai, Y.~Bunimovich, E.~Johnston-Halperin,
  E.~DeIonno, Y.~Luo, B.~A. Sheriff, K.~Xu, Y.~{Shik Shin}, H.-R. Tseng, J.~F.
  Stoddart, and J.~R. Heath.
\newblock {A 160-kilobit molecular electronic memory patterned at 1011 bits per
  square centimetre}.
\newblock {\em Nature}, 445:414, jan 2007.

\bibitem{Jain2018}
M.~Jain, S.~Koren, K.~H. Miga, J.~Quick, A.~C. Rand, T.~A. Sasani, J.~R. Tyson,
  A.~D. Beggs, A.~T. Dilthey, I.~T. Fiddes, S.~Malla, H.~Marriott, T.~Nieto,
  J.~O'Grady, H.~E. Olsen, B.~S. Pedersen, A.~Rhie, H.~Richardson, A.~R.
  Quinlan, T.~P. Snutch, L.~Tee, B.~Paten, A.~M. Phillippy, J.~T. Simpson,
  N.~J. Loman, and M.~Loose.
\newblock {Nanopore sequencing and assembly of a human genome with ultra-long
  reads}.
\newblock {\em Nature Biotechnology}, 36:338, jan 2018.

\bibitem{Jiang2012}
H.~Jiang, M.~D. Riedel, and K.~K. Parhi.
\newblock {Digital Signal Processing With Molecular Reactions}.
\newblock {\em IEEE Design and Test of Computers}, (May/June):21--31, 2012.

\bibitem{Kennedy2018}
E.~Kennedy, P.~Shakya, M.~Ozmen, C.~Rose, and J.~K. Rosenstein.
\newblock {Spatiotemporal information preservation in turbulent vapor plumes}.
\newblock {\em Applied Physics Letters}, 112(26):264103, jun 2018.

\bibitem{Kennedy_metabolites}
{Kennedy et al}.
\newblock Encoding information in synthetic metabolomes.
\newblock {\em PLoS One}, 2019.

\bibitem{PubChem}
S.~Kim, P.~A. Thiessen, E.~E. Bolton, J.~Chen, G.~Fu, A.~Gindulyte, L.~Han,
  J.~He, S.~He, B.~A. Shoemaker, J.~Wang, B.~Yu, J.~Zhang, and S.~H. Bryant.
\newblock {PubChem Substance and Compound databases}.
\newblock {\em Nucleic Acids Research}, 44(D1):D1202--D1213, 2016.

\bibitem{Kosuri2014}
S.~Kosuri and G.~M. Church.
\newblock {Large-scale de novo DNA synthesis : technologies and applications}.
\newblock {\em Nature Methods}, 11(5):499--507, 2014.

\bibitem{Laure2016}
C.~Laure, D.~Karamessini, O.~Milenkovic, L.~Charles, and J.-F. Lutz.
\newblock {Coding in 2D: Using Intentional Dispersity to Enhance the
  Information Capacity of Sequence-Coded Polymer Barcodes}.
\newblock {\em Angewandte Chemie International Edition}, 55(36):10722--10725,
  aug 2016.

\bibitem{Liu1990}
Z.~F. Liu, K.~Hashimoto, and A.~Fujishima.
\newblock {Photoelectrochemical information storage using an azobenzene
  derivative}.
\newblock {\em Nature}, 347:658, oct 1990.

\bibitem{Malinakova2015}
H.~C. Malinakova.
\newblock {Recent advances in the discovery and design of multicomponent
  reactions for the generation of small-molecule libraries}.
\newblock {\em Reports in Organic Chemistry}, pages 75--90, 2015.

\bibitem{Mardis2017}
E.~R. Mardis.
\newblock {DNA sequencing technologies: 2006-–2016}.
\newblock {\em Nature Protocols}, 12:213, jan 2017.

\bibitem{Organick2018}
L.~Organick, S.~D. Ang, Y.-J. Chen, R.~Lopez, S.~Yekhanin, K.~Makarychev, M.~Z.
  Racz, G.~Kamath, P.~Gopalan, B.~Nguyen, C.~N. Takahashi, S.~Newman, H.-Y.
  Parker, C.~Rashtchian, K.~Stewart, G.~Gupta, R.~Carlson, J.~Mulligan,
  D.~Carmean, G.~Seelig, L.~Ceze, and K.~Strauss.
\newblock {Random access in large-scale DNA data storage}.
\newblock {\em Nature Biotechnology}, 36:242, feb 2018.

\bibitem{Polyanskiy2010}
Y.~Polyanskiy, H.~V. Poor, and S.~Verdu.
\newblock {Channel Coding Rate in the Finite Blocklength Regime}.
\newblock {\em IEEE Transactions on Information Theory}, 56(5):2307--2359,
  2010.

\bibitem{Portlock2003}
D.~E. Portlock, D.~Naskar, L.~West, R.~Ostaszewski, and J.~J. Chen.
\newblock {Solid-phase synthesis of five-dimensional libraries via a tandem
  Petasis–Ugi multi-component condensation reaction}.
\newblock {\em Tetrahedron Letters}, 44(27):5121--5124, 2003.

\bibitem{Rang2018}
F.~J. Rang, W.~P. Kloosterman, and J.~de~Ridder.
\newblock {From squiggle to basepair: computational approaches for improving
  nanopore sequencing read accuracy}.
\newblock {\em Genome Biology}, 19(1):90, 2018.

\bibitem{Rose2018}
C.~Rose, S.~Reda, B.~Rubenstein, and J.~Rosenstein.
\newblock {Computing with Chemicals: Perceptrons Using Mixtures of Small
  Molecules}.
\newblock In {\em 2018 IEEE International Symposium on Information Theory
  (ISIT)}, pages 2236--2240, jun 2018.

\bibitem{Ruddigkeit2013}
L.~Ruddigkeit, L.~C. Blum, and J.-L. Reymond.
\newblock {Visualization and Virtual Screening of the Chemical Universe
  Database GDB-17}.
\newblock {\em Journal of Chemical Information and Modeling}, 53(1):56--65, jan
  2013.

\bibitem{Sarkar2016}
T.~Sarkar, K.~Selvakumar, L.~Motiei, and D.~Margulies.
\newblock {Message in a molecule}.
\newblock {\em Nature Communications}, 7:11374, may 2016.

\bibitem{Schena1995}
M.~Schena, D.~Shalon, R.~W. Davis, and P.~O. Brown.
\newblock {Quantitative Monitoring of Gene Expression Patterns with a
  Complementary DNA Microarray}.
\newblock {\em Science}, 270(5235):467 LP -- 470, oct 1995.

\bibitem{Schirwitz2012}
C.~Schirwitz, F.~F. Loeffler, T.~Felgenhauer, V.~Stadler, F.~Breitling, and
  F.~R. Bischoff.
\newblock {Sensing Immune Responses with Customized Peptide Microarrays}.
\newblock {\em Biointerphases}, 7(1):47, 2012.

\bibitem{Schliesser2012}
A.~Schliesser, N.~Picqu{\'{e}}, and T.~W. H{\"{a}}nsch.
\newblock {Mid-infrared frequency combs}.
\newblock {\em Nature Photonics}, 6:440, jun 2012.

\bibitem{Schreiber2000}
S.~L. Schreiber.
\newblock {Target-Oriented and Diversity-Oriented Organic Synthesis in Drug
  Discovery}.
\newblock {\em Science}, 287(5460):1964--1969, 2000.

\bibitem{Soloveichik2010}
D.~Soloveichik, G.~Seelig, and E.~Winfree.
\newblock {DNA as a universal substrate for chemical kinetics}.
\newblock {\em Proceedings of the National Academy of Sciences},
  107(12):5393--5398, 2010.

\bibitem{Tan1999}
D.~S. Tan, M.~A. Foley, B.~R. Stockwell, M.~D. Shair, and S.~L. Schreiber.
\newblock {Synthesis and Preliminary Evaluation of a Library of Polycyclic
  Small Molecules for Use in Chemical Genetic Assays}.
\newblock {\em Journal of the American Chemical Society}, 121(39):9073--9087,
  oct 1999.

\bibitem{Thomas2009}
S.~W. Thomas, R.~C. Chiechi, C.~N. LaFratta, M.~R. Webb, A.~Lee, B.~J. Wiley,
  M.~R. Zakin, D.~R. Walt, and G.~M. Whitesides.
\newblock {Infochemistry and infofuses for the chemical storage and
  transmission of coded information}.
\newblock {\em Proceedings of the National Academy of Sciences}, 106(23):9147
  LP -- 9150, jun 2009.

\bibitem{ibex}
{Unknown artist}.
\newblock {Ibex or Gazelle, Block Print}, 13th or 14th century Egyptian,
  Metropolitan Museum of Art.

\bibitem{amazonomachy}
{Unknown artist}.
\newblock {Terracotta volute-krate}, ca. 450 B.C. Greece, Metropolitan Museum
  of Art.

\bibitem{Way2014}
J.~Way, J.~Collins, J.~Keasling, and P.~Silver.
\newblock {Integrating Biological Redesign: Where Synthetic Biology Came From
  and Where It Needs to Go}.
\newblock {\em Cell}, 157(1):151--161, 2014.

\bibitem{MS_metabolites_database}
{Wishart, D. S. et al.}
\newblock {\em Nucleic Acids Res.}

\bibitem{Zhirnov2016}
V.~Zhirnov, R.~M. Zadegan, G.~S. Sandhu, G.~M. Church, and W.~L. Hughes.
\newblock {Nucleic acid memory}.
\newblock {\em Nature Materials}, 15:366, mar 2016.

\end{thebibliography}

\bibliographystyle{abbrv}

\end{document}